\documentclass[prc,twocolumn,showpacs,showkeys,nofootinbib,superscriptaddress]{revtex4}
\usepackage{graphicx}
\usepackage{dcolumn}
\usepackage{bm}

\begin{document}

\topmargin -0.50in

\title{The $0\nu\beta\beta$-decay nuclear matrix elements 
with self-consistent short-range correlations}

\author{Fedor \v Simkovic}
\affiliation{Institute f\"{u}r Theoretische Physik der Universit\"{a}t
T\"{u}bingen, 72076 T\"{u}bingen, Germany}
\affiliation{Bogoliubov Laboratory of Theoretical Physics, JINR,
141 980 Dubna, Moscow region, Russia}
\affiliation{Department of Nuclear
Physics, Comenius University, Mlynsk\'a dolina F1, 842 15
Bratislava, Slovakia}
\author{Amand Faessler}
\affiliation{Institute f\"{u}r Theoretische Physik der Universit\"{a}t
T\"{u}bingen, 72076 T\"{u}bingen, Germany}
\author{Herbert M\"uther}
\affiliation{Institute f\"{u}r Theoretische Physik der Universit\"{a}t
T\"{u}bingen, 72076 T\"{u}bingen, Germany}
\author{Vadim Rodin}
\affiliation{Institute f\"{u}r Theoretische Physik der Universit\"{a}t
T\"{u}bingen, 72076 T\"{u}bingen, Germany}
\author{Markus Stauf}
\affiliation{
School of Physics and Astronomy, University of Manchester
Manchester, M13 9PL, England}

\begin{abstract}
A self-consistent calculation of nuclear matrix elements of 
the neutrinoless double beta decays ($0\nu\beta\beta$) 
of $^{76}Ge$,  $^{82}Se$,  
$^{96}Zr$,  $^{100}Mo$,  $^{116}Cd$, $^{128}Te$,  $^{130}Te$
and $^{130}Xe$ is presented in the framework of the renormalized 
quasiparticle random phase approximation (RQRPA) and the 
standard QRPA. The pairing and residual interactions 
as well as the two-nucleon short-range correlations are 
for the first time derived  from the same
modern realistic nucleon-nucleon potentials, namely from 
charge-dependent Bonn potential (CD-Bonn) and the Argonne V18
potential. 
In a comparison with the traditional approach of using 
the Miller-Spencer Jastrow correlations matrix elements for the
$0\nu\beta\beta$ decay are obtained, which are larger in magnitude. 
We analyze the differences among
various two-nucleon correlations including those of the unitary
correlation operator method (UCOM) and quantify the uncertainties in
the calculated $0\nu\beta\beta$-decay matrix elements.   
\end{abstract}



\date{\today}

\maketitle

\section{Introduction}

The present data on neutrino oscillations shows that 
pattern of neutrino masses and mixing 
(Pontecorvo-Maki-Nakagava-Sakata mixing matrix)
is not an analogue of that for quarks 
(Cabibbo-Kobayashi-Maskawa quark mixing matrix) 
\cite{Fogli06,Malt08}.
The generation of neutrino masses can be explored,
if the absolute scale of neutrino masses will be
fixed and the issue of leptonic CP violation will be
understood \cite{bil03}. This might  happen, if the 
lepton number violating neutrinoless double beta decay
($0\nu\beta\beta$-decay) will be undoubtable 
observed in running \cite{nemo,gulia}
or planned \cite{gulia,gerda,AEE07,barr}  $0\nu\beta\beta$-decay 
experiments. 

The $0\nu\beta\beta$-decay is a very sensitive probe 
for the Majorana neutrino mass \cite{AEE07,FS98,V02,EV2002,EE04}. 
The $0\nu\beta\beta$-decay 
can occur through different processes but all of them 
require that the neutrino has nonzero mass and is 
a Majorana particle  \cite{schv}. The most proximate or discussed 
theoretical model is to mediate the $0\nu\beta\beta$-decay 
by the exchange of a light Majorana neutrinos. 
A measurement of the decay rate, when combined with neutrino 
oscillation data and a reliable calculation of nuclear 
matrix elements (NMEs), would yield insight into all 
three neutrino mass eigenstates, the type of neutrino mass spectrum
(normal hierarchy or inverted hierarchy) and possibly 
Majorana CP-violating phases.

An important subject in neutrino physics is a reliable calculation
of the $0\nu\beta\beta$-decay NME $M^{0\nu}$ \cite{zuber}. Unfortunately, there are 
no observables that could be directly related to magnitudes of NMEs.
The most popular nuclear structure methods, which has been applied 
for this task, are proton-neutron Quasiparticle Random Phase
Approximation (QRPA) with its variants \cite{Rod06} and the Large-Scale
Shell Model (LSSM) \cite{LSSM1,LSSM2,LSSM3}. 
Recently, there has been significant progress
towards the reduction of uncertainty in the calculated NMEs \cite{Rod06,Rod03a}.
A detailed anatomy of the $0\nu\beta\beta$-decay NMEs pointed out a
qualitative agreement between results of the QRPA-like and LSSM 
approaches \cite{anatomy,smanat}.
It particular, it was shown that only internucleon
distances $r_{ij} \lesssim 2-3~fm$ contribute to $M^{0\nu}$, what
explains a small spread of results for different nuclei. Further, 
it has been shown that correlated NME uncertainties play an important 
role in the comparison of $0\nu\beta\beta$-decay rates for different nuclei
\cite{lisi}.

The improvement of the  calculation of the nuclear matrix elements 
is a very  important and challenging problem. The problem of the
two-nucleon Short-Range Correlations (SRC) have recently inspired  
new $0\nu\beta\beta$-decay studies \cite{Kort07,anatomy,smanat}. 
In the majority of previous calculations
SRC have been treated in a conventional way via Jastrow-type correlation
function in the parametrization of Miller and Spencer \cite{Spencer}. 
Recently, it was found that the consideration of the unitary correlation 
operator method (UCOM)  
leads to increase of the $0\nu\beta\beta$-decay NME by about 20-30\%
\cite{Kort07,anatomy,smanat}.
It was concluded that we do not know the best way to treat the SRC,
a fact that contributes to uncertainties. 

In the present article we improve on the Miller-Spencer
Jastrow and the UCOM 
and perform a selfconsistent calculation of the 
$0\nu\beta\beta$-decay NMEs by considering pairing,
ground state and short-range correlations deduced from the
same realistic nucleon-nucleon (NN) interaction. In particular,
the two-nucleon short-range correlations will be determined 
within the ${\it coupled~ cluster}$ or $exponential~ S$ approach 
by using CD-Bonn and Argonne V18 NN-forces \cite{muether,stauf} 
and compared with Jastrow and UCOM SRC.
Then, they will be used in the QRPA and renormalized QRPA
(RQRPA) calculations 
of the $0\nu\beta\beta$-decay NMEs of experimental interest.

The paper is organized as follows:
In Sec. II the formalism of the $0\nu\beta\beta$-decay associated 
with exchange of light Majorana neutrinos is briefly
reviewed. Sec. II is devoted to the analysis of different 
treatments of the two-nucleon short-range correlations in the
context of the correlated $0\nu\beta\beta$-decay operator.
In Sec. IV we present numerical results for nuclei
of experimental interest. Section V summarizes our findings. 

\section{Formalism}

In this section we present basic expressions associated with 
the calculation of the $0\nu\beta\beta$-decay NME, what allow 
us to discuss the effects of Finite Nucleon Size  (FNS) 
and two-nucleon SRC.

By assuming the dominance of the light neutrino mixing 
mechanism the inverse value of the $0\nu\beta\beta$-decay 
half-life for a given isotope $(A,Z)$ is given by 
\begin{equation}
\frac{1}{T_{1/2}} = G^{0\nu}(E_0,Z) |{M'}^{0\nu}|^2
|\langle m_{\beta\beta} \rangle|^2.
\end{equation}
Here, $G^{0\nu}(E_0,Z)$ and ${M'}^{0\nu}$ are, respectively,
the known phase-space factor ($E_0$ is the energy release)
and the nuclear matrix element, which depends 
on the nuclear structure of the particular isotopes 
$(A,Z)$, $(A,Z+1)$ and $(A,Z+2)$ under study. Under the 
assumption of the mixing of three light massive Majorana 
neutrinos the effective Majorana neutrino mass 
$\langle m_{\beta\beta} \rangle$ takes the form
\begin{equation}
\langle m_{\beta\beta} \rangle = \sum_i^N |U_{ei}|^2 e^{i\alpha_i} m_i ~,
~({\rm all~} m_i \ge 0)~,
\end{equation}
where $U_{ei}$ is the first row of the neutrino mixing matrix and
the and $\alpha_i$ are unknown Majorana phases. It is apparent 
that any uncertainty in ${M'}^{0\nu}$ makes the value of 
$\langle m_{\beta\beta} \rangle$ equally uncertain.

Our phase space factors $G^{0\nu}(E_0,Z)$, which include fourth
power of axial-coupling constant $g_A = 1.25$, are tabulated in Ref.\
\cite{Sim99}. They agree quite closely with those given earlier in
Ref. \cite{BV92}.  The $G^{0\nu}(E_0,Z)$ contain the inverse
square of the nuclear radius $(R_{nucl})^{-2}$, compensated by the
factor $R_{nucl}$ in ${M'}^{0\nu}$. Different authors 
use different conventions for $R_{nucl}$ ($R_{nucl}= r_0 A^{1/3}$
$r_{0} = 1.2~ fm$ or $r_0 = 1.1~fm$), a fact that is important 
to keep in mind when comparing the matrix elements without 
also looking at $G^{0\nu}(E_0,Z)$.

The nuclear matrix element ${M'}^{0\nu}$ is defined as 
\begin{equation}
{M'}^{0\nu} =  \left(\frac{g_A}{1.25}\right)^2 {M}^{0\nu},
\end{equation}
where ${M}^{0\nu}$ consists of Fermi, Gamow-Teller and tensor parts as
\begin{equation}
{M}^{0\nu} =  - \frac{M_{F}}{g^2_A} + M_{GT} + M_T.
\end{equation}
This definition of ${M'}^{0\nu}$ \cite{Rod06} allows
to display the effects of uncertainties in $g_A$ and to use
the same phase factor $G^{0\nu}(E_0,Z)$ when calculating 
the $0\nu\beta\beta$-decay rate.

In the QRPA (and RQRPA) ${M}^{0\nu}$ is written as sums 
over the virtual
intermediate states, labeled by their angular momentum and parity
$J^{\pi}$ and indices $k_i$ and $k_f$ 
\begin{eqnarray}
M_K  =  \sum_{J^{\pi},k_i,k_f,\mathcal{J}} \sum_{pnp'n'}
(-1)^{j_n + j_{p'} + J + {\mathcal J}} \times~~~~~~~~~~
\nonumber\\
\sqrt{ 2 {\mathcal J} + 1}
\left\{
\begin{array}{c c c}
j_p & j_n & J  \\
 j_{n'} & j_{p'} & {\mathcal J}
\end{array}
\right\}  \times~~~~~~~~~~~~~~~~~~~
\nonumber \\
\langle p(1), p'(2); {\mathcal J} \parallel 
{\cal O}_K 
\parallel n(1), n'(2); {\mathcal J} \rangle \times~~
\nonumber \\
\langle 0_f^+ ||
[ \widetilde{c_{p'}^+ \tilde{c}_{n'}}]_J || J^{\pi} k_f \rangle
\langle  J^{\pi} k_f |  J^{\pi} k_i \rangle
 \langle  J^{\pi} k_fi|| [c_p^+ \tilde{c}_n]_J || 0_i^+ \rangle ~.
\nonumber\\
\label{eq:long}
\end{eqnarray}
The reduced matrix elements of the one-body operators
$c_p^+ \tilde{c}_n$ ($\tilde{c}_n$ denotes the time-reversed state)
in the Eq. (\ref{eq:long})
depend on the BCS coefficients $u_i,v_j$ and on the QRPA vectors
$X,Y$ \cite{Sim99}. The difference between QRPA and RQRPA resides
in the way these reduced matrix elements are calculated.

The two-body operators $O_K, K$ = Fermi (F), Gamow-Teller (GT), and Tensor
(T) in (\ref{eq:long}) contain neutrino potentials and spin and isospin 
operators, and RPA energies $E^{k_i,k_f}_{J^\pi}$: 
\begin{eqnarray}
O_F(r_{12},E^k_{J^\pi}) &=& \tau^+(1)\tau^+(2)~ H_F(r_{12},E^k_{J^\pi})~,
\nonumber\\
O_{GT}(r_{12},E^k_{J^\pi}) &=& \tau^+(1)\tau^+(2)~ H_{GT}(r_{12},E^k_{J^\pi})
~\sigma_{12}~,
\nonumber\\
O_{T}(r_{12},E^k_{J^\pi}) &=& \tau^+(1)\tau^+(2)~ H_{T}(r_{12},E^k_{J^\pi})
~S_{12}.
\label{a12}
\end{eqnarray}
with
\begin{eqnarray}
{\bf r}_{12} &= &{\bf r}_1-{\bf r}_2, ~~~ r_{12} = |{\bf r}_{12}|,
~~~
\hat{{\bf r}}_{12} = \frac{{\bf r}_{12}}{r_{12}},~\nonumber \\
\sigma_{12} &=& {\vec{ \sigma}}_1\cdot {\vec{ \sigma}}_2 \nonumber\\
S_{12} &=& 3({\vec{ \sigma}}_1\cdot \hat{{\bf r}}_{12})
       ({\vec{\sigma}}_2 \cdot \hat{{\bf r}}_{12})
      - \sigma_{12}.
\end{eqnarray}
Here, ${\bf r}_1$ and ${\bf r}_2$ are the coordinates of the
nucleons undergoing beta decay.

The neutrino potentials are integrals over the exchanged momentum $q$,
\begin{eqnarray}
H_K (r_{12},E^k_{J^\pi}) =~~~~~~~~~~~~~~~~~\nonumber\\
\frac{2}{\pi} {R} \int_0^{\infty}~ f_K(qr_{12})~
\frac{ h_K (q^2) q dq }
{q + E^k_{J^\pi} - (E_i + E_f)/2} ~,
\label{eq:pot}
\end{eqnarray}
The functions $f_{F,GT}(qr_{12}) = j_0(qr_{12})$ and $f_{T}(qr_{12})
= - j_2(qr_{12})$ are spherical Bessel functions.

The potentials (\ref{eq:pot}) depend explicitly, though rather 
weakly, on the energies of the virtual intermediate states, 
$E^k_{J^\pi}$. 
The functions $h_K(q^2)$ that enter the $H_K$'s 
through the integrals over $q$ in Eq.\ (\ref{eq:pot}) are 
\begin{eqnarray}
h_{F} ({ q}^{2})  & = & g^2_V({ q}^{2}) \nonumber \\
h_{GT} ({ q}^{2}) & = & \frac{g^2_A({ q}^{2})}{g^2_A} 
[ 1 - \frac{2}{3} \frac{ { q}^{2}}{ { q}^{2} + m^2_\pi } + 
\frac{1}{3} ( \frac{ { q}^{2}}{ { q}^{2} + m^2_\pi } )^2 ]
\nonumber\\
&& + \frac{2}{3} \frac{g^2_M({ q}^{2} )}{g^2_A} \frac{{ q}^{2} }{4 m^2_p }, 
\nonumber \\
h_T ({ q}^{2}) & = & \frac{g^2_A({ q}^{2})}{g^2_A} [ 
\frac{2}{3} \frac{ { q}^{2}}{ { q}^{2} + m^2_\pi } -
\frac{1}{3} ( \frac{ { q}^{2}}{ { q}^{2} + m^2_\pi } )^2 ]
\nonumber\\
&& + \frac{1}{3} \frac{g^2_M ({ q}^{2} )}{g^2_A} \frac{{ q}^{2} }{4 m^2_p }    
\end{eqnarray}
Here, we used the Partially Conserved Axial Current (PCAC) hypothesis.

The FNS is taken into account via momentum dependence 
of the nucleon form-factors. 
For the vector, weak-magnetism and axial-vector form factors we adopt 
the usual dipole approximation as follows: 
\begin{eqnarray}
g_V({ q}^{2}) &=& \frac{g_V}{(1+{ q}^{2}/{M^2_V})^2}, 
\nonumber\\
g_M({ q}^{2}) &=& (\mu_p-\mu_n) g_V({ q}^{2}), 
\nonumber\\
g_A({ q}^{2}) &=& \frac{g_A}{(1+{ q}^{2}/{M^2_A})^2},
\end{eqnarray}
where $g_V = 1$, $g_A = 1.00$ (quenched) and $1.25$ (unquenched), 
$(\mu_p - \mu_n) = 3.70$. The parameters $M_V = 850 ~MeV$ and 
$M_A = 1~086 ~MeV$ come from electron scattering and 
neutrino charged-current scattering experiments.

The $0\nu\beta\beta$-decay matrix elements were usually calculated in
some approximations, which are only partially justified 
(see also discussion in \cite{Rod06}):\\
i) The effect of higher order terms of nucleon currents 
was not taken into account. In this case we have
\begin{equation}
h_{F} ({ q}^{2}) =  g^2_V({ q}^{2}), ~~~~~~~
h_{GT} ({ q}^{2}) = \frac{g^2_A({ q}^{2})}{g^2_A},~~~~~ 
h_T ({ q}^{2}) = 0.
\end{equation}
We note that if in addition nucleons are considered to be point-like 
$h_F$ and $h_{GT}$ are equal to unity.\\
ii) The closure approximation for intermediate nuclear states was considered
by replacing energies of intermediate states $[E^k_{J^\pi} - (E_i + E_f)/2]$ 
by an average value ${\overline E} \approx 10~MeV$.\\
Within these approximations the neutrino potential in Eq. (\ref{eq:pot})
can be written as \cite{tomoda}
\begin{eqnarray}
H_{bare}(r_{12}, {\overline E}) &=& \frac{2}{\pi}
\left[\sin{({\overline E} r_{12})} Ci{({\overline E} r_{12})} \right. \nonumber\\
&-& \left.  \cos{({\overline E} r_{12})} Si{({\overline E} r_{12})}
\right] \frac{R_{nucl}}{r_{12}}.\nonumber\\
\end{eqnarray}
Here, $Ci(x)$ and $Si(x)$ are the cosine and sine integrals, respectively.
The value of ${\overline E}$ has practically no impact on the behavior 
of neutrino exchange potential at short internucleon distances.  
In the limit ${\overline E}=0$ and zero neutrino mass the neutrino potential 
is coulombic: $H_{bare}(r_{12},{\overline E}=0) = R_{nucl}/r_{12}$. 

It is worth to mention some general properties of the Fermi $M_F$ and 
the Gamow-Teller $M_{GT}$ matrix elements, in particular 
some multipole contributions of states of the intermediate 
odd-odd nucleus are equal to zero. We have
\begin{eqnarray}
M_F(J^+) &=& 0~~~~ for~odd~J, \nonumber\\
M_F(J^-) &=& 0~~~~ for~even~J, \nonumber\\
M_{GT}(0^+) &=& 0.  
\label{condzero}
\end{eqnarray} 

\section{Short range correlations for the $0\nu\beta\beta$-decay}

An important component of the $M_K$ in (\ref{eq:long}) 
is an unantisymmetrized two-body matrix element, 
\begin{equation}
\langle p(1), p'(2); {\mathcal J} \parallel 
{\cal O}_K 
\parallel n(1), n'(2); {\mathcal J} \rangle, 
\end{equation}
constructed from two one-body matrix elements by coupling pairs of
protons and neutrons to angular momentum ${\cal J}$. We note, 
in the closure approximation, i.e., if energies of intermediate states 
$(E^k_{J^\pi} - E_i)$ are replaced by an average value ${\overline E}$, 
and the sum over intermediate states is taken by closure, 
$\sum_n |J^\pi_k><J^\pi_k| = 1$, we end up with antisymmetrized
two-body matrix elements.  As the virtual neutrino has an average 
momentum of  $\sim 100$ MeV \cite{anatomy}, 
considerably larger than the differences 
in nuclear excitation, the closure approximation limit is found
to be meaningful showing on the importance of correlations 
of the two $\beta$-decaying nucleons.

\subsection{The Jastrow and UCOM short-range correlations}

The QRPA (RQRPA) as well as the LSSM
approaches do not allow to introduce short-range 
correlations into the two-nucleon relative wave function. The 
traditional way is to introduce an explicit Jastrow-type correlation 
function $f(r_{12})$ into the involved two-body transition 
matrix elements
\begin{eqnarray}
 \langle {{\Psi}}_{\mathcal J} 
\parallel f(r_{12}) 
{\cal O}_K (r_{12}) f(r_{12}) 
\parallel {{\Psi}}_{\mathcal J}  \rangle 
~~~~~~~\nonumber\\
\equiv ~ \langle {\overline{\Psi}}_{\mathcal J} 
\parallel {\cal O}_K (r_{12}) 
\parallel {\overline{\Psi}}_{\mathcal J}  \rangle.
~~~~~~~~~~~~~~
\label{correl}
\end{eqnarray}
Here, 
\begin{eqnarray}
|{\overline{\Psi}}_{\mathcal J}  \rangle
&=&  f(r_{12}) ~| {{\Psi}}_{\mathcal J} \rangle,   
\nonumber\\
| {{\Psi}}_{\mathcal J}  \rangle &\equiv&
| n(1), n'(2); {\mathcal J} \rangle 
\label{cwf}
\end{eqnarray}
are the relative wave function with and without the short-range
correlations, respectively.
In the parametrization of Miller and Spencer 
\cite{Spencer} we have
\begin{equation}
f(r_{12}) = 1 - e^{-a r^2}(1-b r^2), ~a=1.1~fm^{-2},~ b=0.68~fm^{-2}.
\end{equation}
These two parameters ($a$ and $b$) 
are correlated and chosen in the way that the norm 
of the relative wave function 
$|{\overline{\Psi}}_{\mathcal J}  \rangle$ is conserved. 

Usually, the nuclear matrix element $M^{0\nu}$ is calculated 
in relative and center-of-mass coordinates as the Jastrow
correlation function depends only on $r_{12}$. This
is achieved with help of the well-known Talmi-Moshinski 
transformation \cite{talmi} for the harmonic oscillator basis. 
Within this procedure the chosen construction of the relative
wave function,  namely a product of $f(r_{12})$ with 
harmonic oscillator wave function in (\ref{cwf}), 
is well justified. Any more  complex structure of correlation
function, e.g., a consideration of different correlation functions 
for different channels, would result in violation of requirements 
in (\ref{condzero}) as the  Talmi-Moshinski transformation 
is considered.

Recently, it was proposed \cite{Kort07} to adopt instead of the Jastrow method 
the UCOM approach for description 
of the two-body correlated wave function \cite{UCOM}. This approach 
describes not only short-range, but also central and tensor correlations 
explicitly by means of a unitary transformation. 
The state-independent short-range  correlations  
are treated explicitly  while long-range correlations should 
be described in a model space. Applied to a realistic 
NN interaction, the method produces a correlated interaction, 
which can be used as a universal effective interaction, 
for calculations within simple Hilbert spaces.
The UCOM method produces good results for 
the binding energies of nuclei already at the Hartree-Fock 
level \cite{roth}. There are also some first applications
for description of collective multipole excitations \cite{paar}.

Within the UCOM the short-range and long-range
correlations are imprinted into uncorrelated many 
body states by a unitary transformation. In the case 
of the $0\nu\beta\beta$-decay calculation the
correlated two-nucleon wave function was taken 
as \begin{equation}
| {{\Psi}}_{\mathcal J} \rangle
=
C_r | {{\Psi}}_{\mathcal J} \rangle.
\end{equation}  
Here, $C_r$ is the unitary correlation operator
describing the short-range correlations. The explicit 
form of $C_r$ is given in \cite{UCOM} with a separate 
parametrization for different LS-channels. 
In application to the $0\nu\beta\beta$-decay this fact 
leads to a slight violation of conditions  (\ref{condzero}) 
when Talmi-Moshinski transformations are considered. 
The UCOM-corrected NMEs of the $0\nu\beta\beta$-decay are
significantly less suppressed when compared with
those calculated with Jastrow SRC \cite{anatomy,Kort07}.

\subsection{Self-consistent two-body short-range correlations}

The two-nucleon wave function with short range correlations 
can be calculated from the same realistic NN-interaction, 
which is used in the derivation of the Brueckner G-matrix 
elements of the nuclear Hamiltonian. A method of choice is,
e.g., the Brueckner-Bethe hole-line expansion, the coupled cluster
method (CCM) or exponential S approach and the approach of 
self-consistent evaluation of Green functions \cite{muether}. 

There are various  modern NN potentials, which yield a very accurate 
fit to the nucleon-nucleon scattering phase shifts. Two of them 
are the so-called charge-dependent Bonn potential (CD-Bonn) 
\cite{cdbonn} and the Argonne V18 potential (Argonne) \cite{argonne}.
They differ in description of both long-range
and short-range parts of NN-interaction. The CD-Bonn is derived
in the framework of the relativistic meson field theory. The Argonne
potential is a purely local potential in the sense that it uses
the local form of the one-pion exchange potential for the long-range
part and parametrizes the contributions of medium and short-range
distances in terms of local functions multiplied by a set of spin-isospin
operators.

We have chosen  the CCM \cite{stauf} 
to evaluate the effect of short-range correlation on 
the $0\nu\beta\beta$-decay NMEs, because it provides 
directly correlated two-body wave functions.
The basic features of the CCM are described in the review
article by K\"ummel \cite{ccm2}. The  developments of this
many-body approach with applications can be found in \cite{ccm3,ccm4}.

The CCM starts by assuming an appropriate Slater determinant 
$|\Phi \rangle$ as a first approximation for the exact eigenstate 
of the A-particle system. The many-body wave function of the
coupled cluster or exp(S) method can be written as
\begin{equation}
| \Phi \rangle_{corr.}  = 
exp\left(\sum^A_{n=1} {\hat S}_n\right) | \Phi \rangle.
\label{eq:ccmansatz}
\end{equation}
The n-particle n-hole excitation operator ${\hat S}_n$ is
given by
\begin{eqnarray}
{\hat S}_n = \frac{1}{(n!)^2}\times
~~~~~~~~~~~~~~~~~~~~~~~~ \nonumber\\
\sum_{\nu_i,\rho_i} \langle \rho_1 \dots \rho_n |S_n|
\nu_1\dots \nu_n\rangle
 a^\dagger_{\rho_1}\dots a^\dagger_{\rho_n}
 a_{\nu_n}\cdots a_{\nu_1}.
\label{Sop}
\end{eqnarray}
The sum in (\ref{Sop}) is restricted to states
$\rho_i$ which are unoccupied in the model state $|\Phi \rangle$,
while states $\nu_i$ refer to states which are occupied in
$|\Phi \rangle$.

\vspace{0.8cm}
\begin{figure}[tb]
\includegraphics[width=.47\textwidth,angle=0]{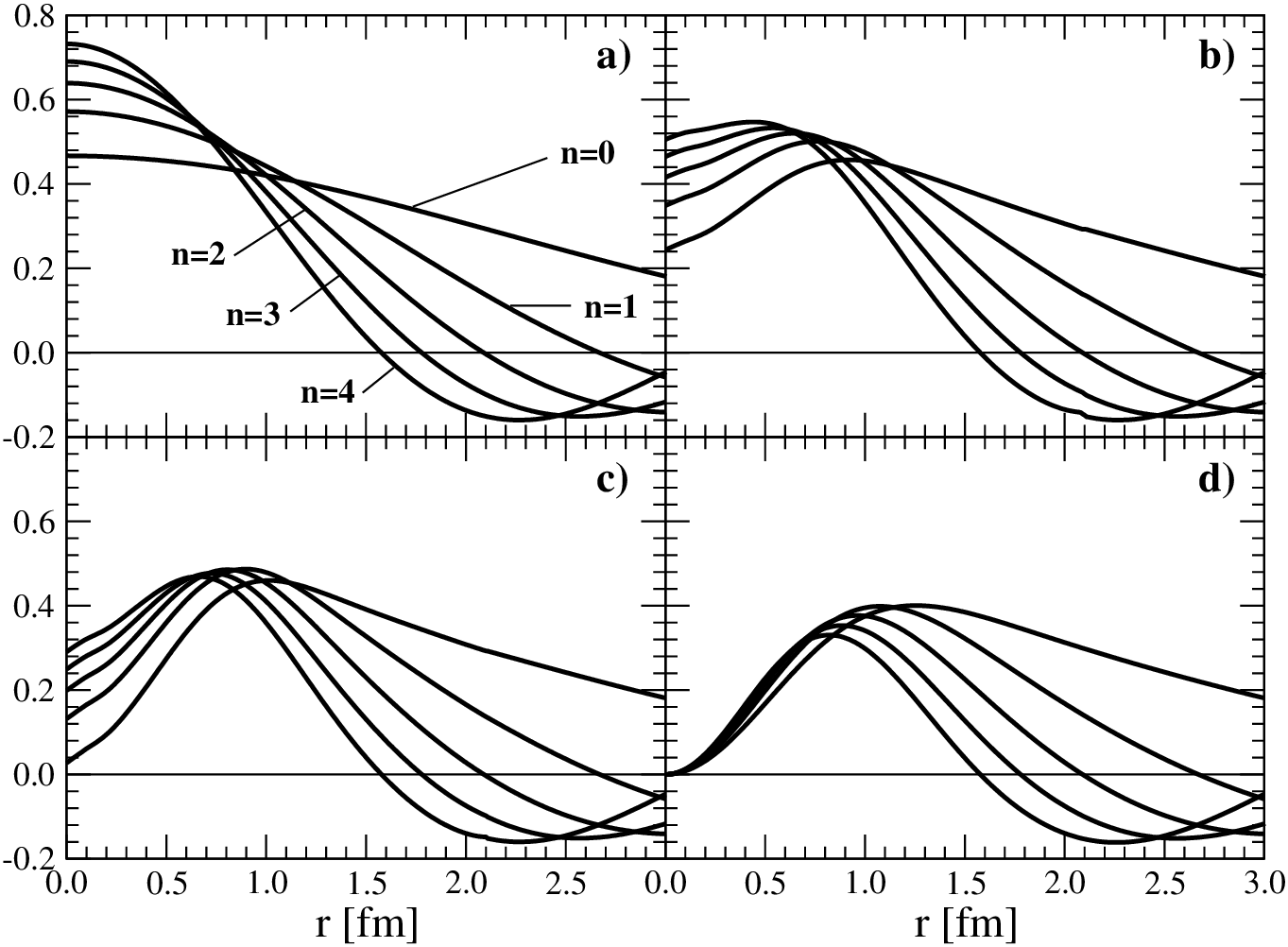}
\caption{Two nucleon wave functions 
as a function of the relative distance for the $^{1}S_0$
partial wave and radial quantum numbers $n = 0,1,2,3$ and 
$4$.  The results in panel a) are for uncorrelated 
two-nucleon wave functions. The results in panels b), c) and d),
respectively, are with coupled cluster method with CD-Bonn potential, 
coupled cluster method with Argonne potential
and Miller-Spencer Jastrow short-range correlations. The 
harmonic oscillator parameter b is $2.18~fm$. 
} 
\label{fig:cowf}
\end{figure}
 
A Slater determinant of harmonic oscillator wave functions 
is considered for $|\Phi \rangle$. For a given nuclear
system of interest an appropriate value of the oscillator 
length $b$ is chosen.
In the so-called $S_2$ approximation of the CCM the amplitudes defining $\hat
S_n$ with $n>3$ in (\ref{eq:ccmansatz}) are ignored. This means that effects
beyond Hartree-Fock and two-body correlations (i.e. genuine three- and 
more-particle correlations) are ignored. This leads to a
coupled set of equations for the evaluation of the correlation
operators ${\hat S}_1$ and ${\hat S}_2$ \cite{stauf}. Therefore this $S_2$
approximation corresponds essentially to the Brueckner-Hartree-Fock (BHF) 
approximation of the hole-line expansion or Brueckner theory. In fact, the
hole-hole scattering terms, which are included in the $S_2$ but ignored in BHF
turn out to yield small effects only. Therefore it is consistent to combine
the correlation effects from CCM with the matrix elements of the $G$-matrix,
the effective interaction determined in the BHF approximation.

The use of the oscillator ansatz in Slater determinant $| \Phi>$ in 
eq.(\ref{eq:ccmansatz}) leads to an evaluation of the correlated two-nucleon
wave functions in terms of product
wave functions for the relative and center-of-mass
coordinates. The two-body states take the form
\begin{equation}
| \left[n(lS)j\right]NL {\cal J} \tau\rangle.
\end{equation}
Here, N and L denote the harmonic-oscillator quantum numbers
for the center of mass wave function and $l$ refers to the
orbital angular momentum for the relative motion, which is 
coupled with a total spin of the pair $S$ to angular momentum 
${\cal J}$. The basis states for the radial part of this relative
motion are labeled by a quantum number $n$. 

As an example we present in Fig. \ref{fig:cowf} 
relative wave functions for correlated and uncorrelated 
two-body wave functions in the case of  $^{1}S_0$ partial waves 
and different values of the radial quantum numbers $n$
($n=0,~1,~2,~3$ and $4$). In panel a) uncorrelated harmonic 
oscillator wave functions are plotted. Panels b) and  
c) show the relative wave functions obtained with help
of CCM employing the  CD-Bonn and the Argonne potentials. 
 For a comparison
relative wave functions with Miller-Spencer Jastrow SRC
are displayed in panel d). While the Jastrow ansatz suppresses the relative
wave function in the limit $r_{12}\to 0$ completely, we find that this suppression
effect is much weaker in the CCM calculation. This is true even if the Argonne
potential is used, which is known to produce stronger short-range components
than the softer CD-Bonn potential. Also note that the correlated wave
functions derived from realistic interactions exhibit a short-range behavior
which depends on the radial quantum number $n$, whereas the Jastrow approach
yields almost identical relative wave functions for small values of $r_{12}$.

components of the NN interaction at short distances 
are weaker for the CD-Bonn potential in comparison with 
Argonne interaction. But, in the case of the Jastrow SRC 
the reduction of the relative wave function 
for small values of $r_{12}$ is even much stronger. 

The advantage of the CCM \cite{stauf} 
is a factorization of the correlated two-body wave function
on a product of a correlation function and a harmonic oscillator
wave function. This allows us to discuss the effect of 
the SRC in terms of the correlated operator, which is a
product of the transition operator ${\cal O}_K (r_{12})$ and
two correlation functions $f(r_{12})$ (see Eqs. (\ref{correl})). 
For our purposes we consider CCM CD-Bonn $f_{B}$ 
and CCM Argonne $f_{A}$ correlation functions deduced
from the $^{1}S_0 (n=0)$ correlated two-body wave function.
The use of this single correlation function for all partial 
waves and quantum numbers $n$ is numerically well justified 
and is dictated by the use of the Talmi-Moshinski transformation
in the evaluation of the $0\nu\beta\beta$-decay matrix element.
 
\begin{figure}[tb]
\includegraphics[width=.47\textwidth,angle=0]{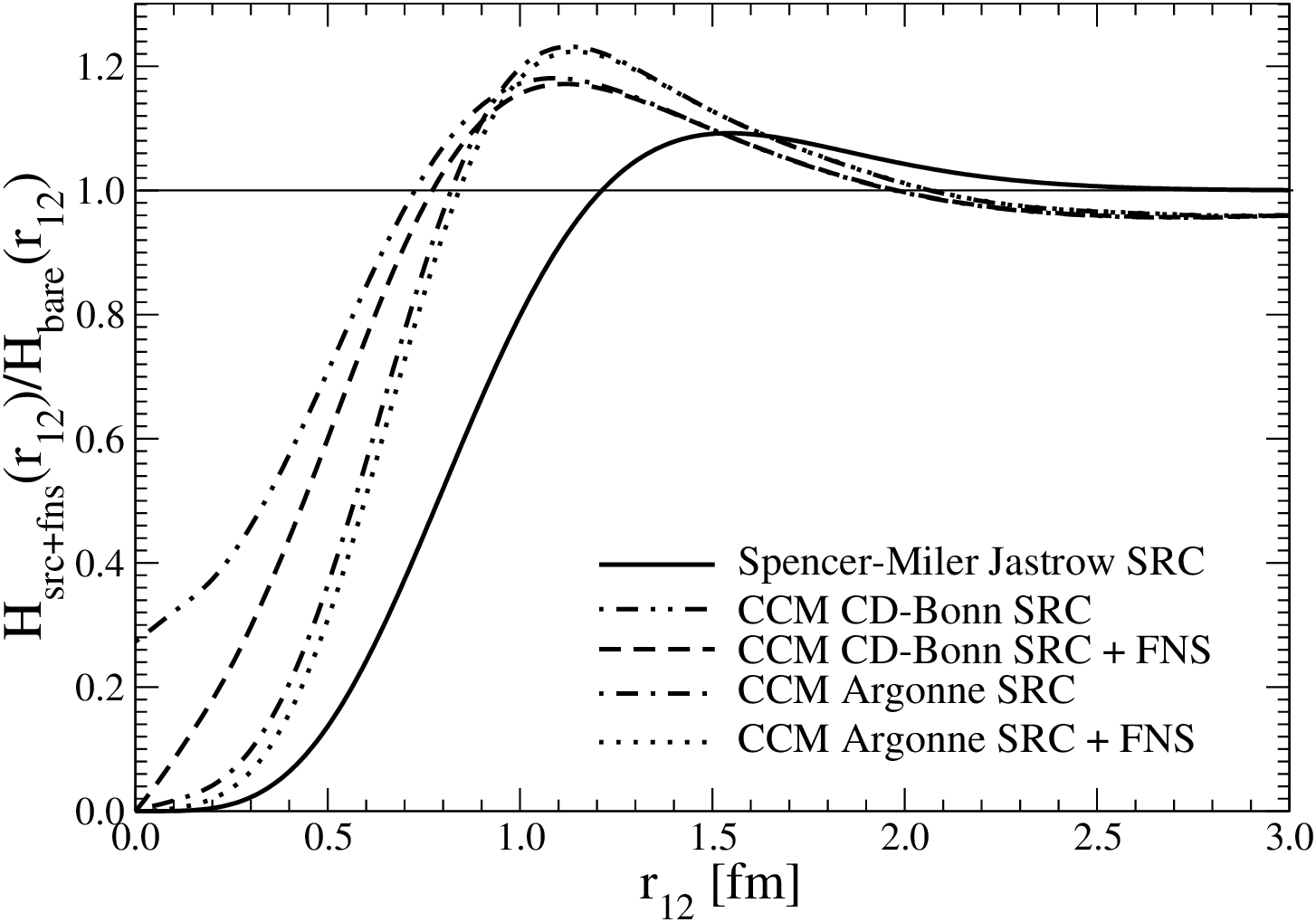}
\caption{The ratio of neutrino potentials with and without two-nucleon 
short-range correlations (SRC). Results are shown for 
the CCM CD-Bonn and Argonne and Miller-Spencer SRC 
with and without consideration of the effect of finite size 
of a nucleon.  It is assumed $\overline{E} = 8~MeV$.
} 
\label{fig:cop}
\end{figure}

In Fig.  \ref{fig:cop} the differences between the CCM 
and the Miller-Spencer SRC are manifested by plotting
the ratio of correlated $H_{src+fns}(r_{12})$ and 
uncorrelated $H_{bare}(r_{12})$ neutrino potentials. 
The averaged energy of intermediate
 nuclear states ${\overline E}$ is 8 MeV. For sake of simplicity 
the effect of higher order terms of nucleon currents on 
the neutrino potential is neglected.
From Fig.  \ref{fig:cop} we see that there is a significant 
difference between the CCM and the Miller-Spencer treatment
of the SRC. The maxima of the CCM and the Spencer-Miller 
curves occur at $1~fm$ and $1.5~fm$, respectively. One finds 
also that the reduction at short distances is much weaker 
for CD-Bonn than for Argonne interactions. 

For purpose of numerical calculation of the $0\nu\beta\beta$-decay
NMEs we present the CCM short-range correlation functions 
in an analytic form of Jastrow-like function as
\begin{equation}
f_{A,B}(r_{12}) = 1 ~-~ c~ e^{-a r^2}(1-b r^2). 
\end{equation}
The set of parameters for Argonne  and CD-Bonn NN interactions 
is given by
\begin{eqnarray}
f_{A}(r_{12}):~~a &=& 1.59~fm^{-2},~~b = 1.45~fm^{-2},~~c = 0.92,\nonumber\\
f_{B}(r_{12}):~~a &=& 1.52~fm^{-2},~~b = 1.88~fm^{-2},~~c = 0.46.\nonumber\\
\end{eqnarray}
The calculated NMEs with these short-range correlation functions agree within 
a few percentages with those obtained without this approximation.
We note that the dependence of the SRC on the value of oscillator length $b$ 
is rather weak.  

\begin{figure}[tb]
\includegraphics[width=.47\textwidth,angle=0]{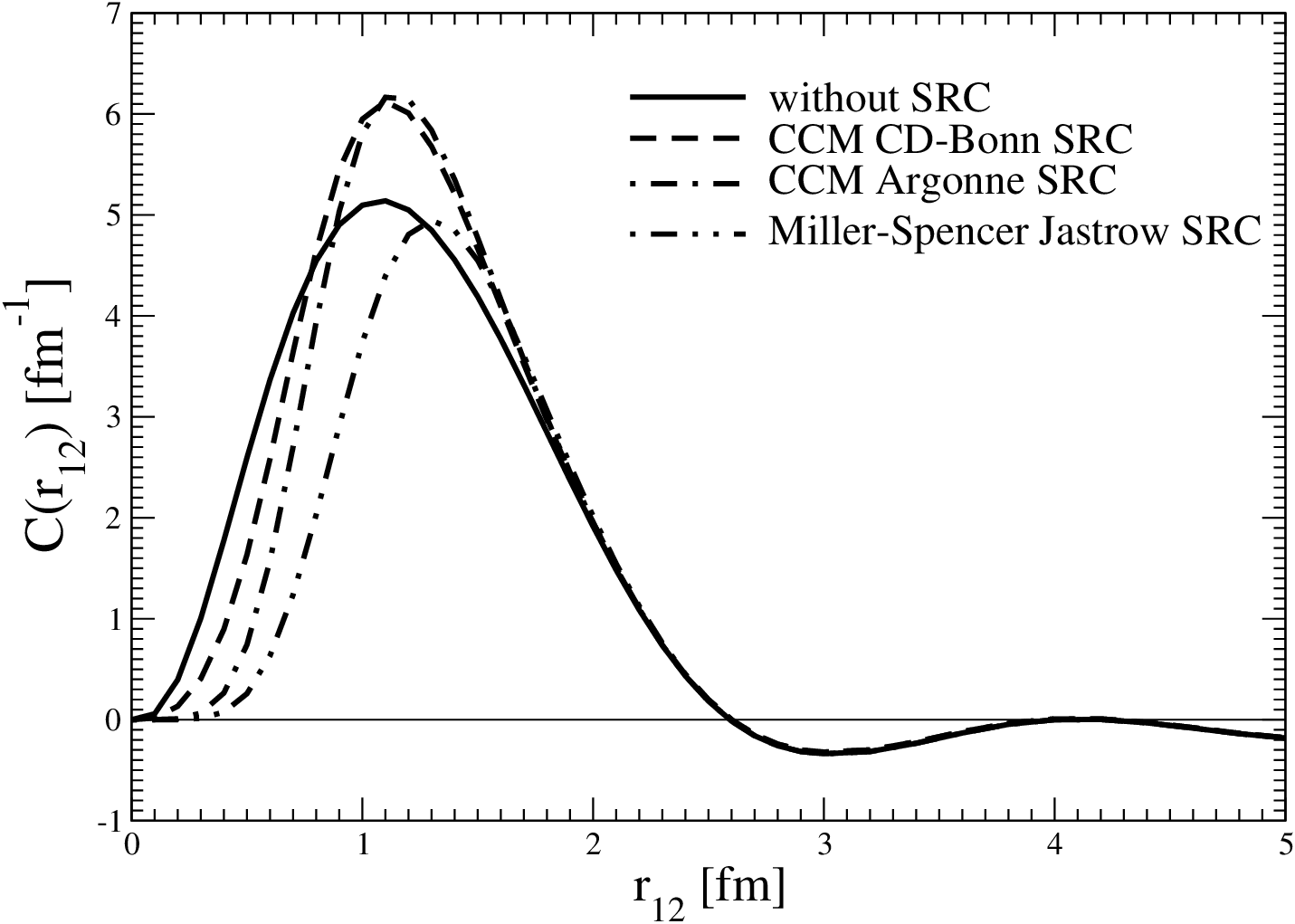}
\caption{The $r_{12}$ dependence of $M^{0\nu}$
in $^{76}Ge$ evaluated in the model space that 
contains twelve subshells. The four curves show
the effects of different treatment of short-range 
correlations (SRC). The finite nucleon size (FNS) 
is taken into account.  
} 
\label{fig:src}
\end{figure}

In Fig. \ref{fig:src} the $r_{12}$ dependence of $M^{0\nu}$ is shown 
for CCM Argonne, CCM CD-Bonn and phenomenological Jastrow
SRC for the $0\nu\beta\beta$-decay of $^{76}Ge$.
The quantity $C(r_{12})$ is defined by 
\begin{equation}
 M^{0\nu} = \int_0^\infty C(r_{12}) dr_{12}. 
\end{equation} 
We note that the range of $r_{12}$ is practically restricted from
above by  $r_{12} \le 2 R_{nucl}$. From the Fig. \ref{fig:src}
we see that a modification of the neutrino potential due to
the different types of SRC is transmitted to the behavior of $C(r_{12})$
for $r_{12} \le 2 fm$. Both, the CCM short-range correlation
functions (see Fig. \ref{fig:cop}) and $C(r_{12})$ with
SRC switched off (but with FNS effect) have maxima for $r_{12}\simeq 1~ fm$
unlike the phenomenological Jastrow function with maximum shifted 
to $r_{12} = 1.5~ fm$. This explains a significant increase 
of $C(r_{12})$ with CCM SRC  and suppression of 
$C(r_{12})$ with Jastrow SRC in this region.
This phenomenon clarifies also why the values of $M^{0\nu}$ 
obtained with  CCM SRC are comparable 
to those calculated when only the FNS effect is considered 
(see Table \ref{tab:fns}). The increase of $C(r_{12})$ 
for $r_{12} \simeq 1~fm$ compensates its reduction in the range 
$r_{12} \le 0.7~fm$.

\subsection{Finite nucleon size and two-body short range correlations}

The FNS effects are introduced in the calculation of the $0\nu\beta\beta$-decay 
NMEs by the dipole form factors in momentum space. 
The form factor simulates the fact that nucleon is not 
a point particle, and therefore as $q^2$ increases, the probability that
nucleon will stay intact (and not produce pions etc) decreases.
The physics of FNS and SRC is different, but both reduce
the magnitude of the operator when $q^2$ increases or equivalently
$r_{12}$ decreases. It was found \cite{anatomy}
that Miller-Spencer and the UCOM short-range correlations essentially 
eliminate the effect of the FNS on the matrix elements. The same is
expected to be valid also for the CCM CD-Bonn and Argonne short-range
correlations. From Fig. \ref{fig:cop} we see that the ratio of 
correlated and uncorrelated neutrino potentials is changed only weakly
if in addition to two-nucleon SRC  the effect of the FNS is taken into account.   

\begin{figure}[tb]
\includegraphics[width=.47\textwidth,angle=0]{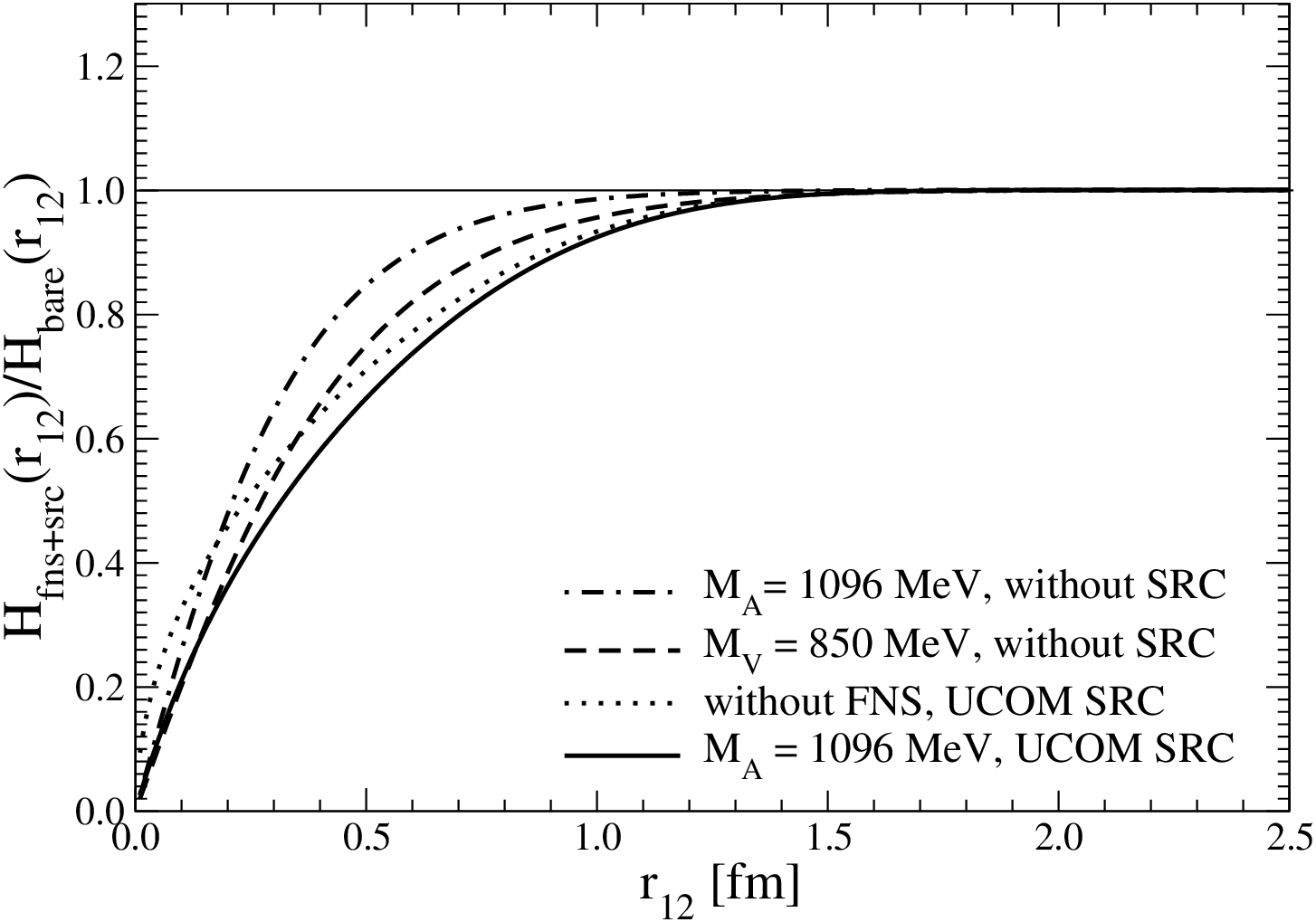}
\caption{A ratio of neutrino potentials with  
inclusion of the finite nucleon size effect (FNS) 
and bare neutrino potential for cut-off masses $M_{V}$ and $M_{V}$. 
This is compared with ratio of neutrino potential with UCOM short 
range correlations (SRC) and bare neutrino potential. $\overline{E} = 8~MeV$
is used in calculation.
} 
\label{fig:fns}
\end{figure}

It is worth mentioning that the behavior of the UCOM  correlated neutrino
potential differs strongly from those calculated with the CCM and Jastrow 
SRC. This is manifested
in Fig. \ref{fig:fns}. The studied ratio of UCOM correlated and uncorrelated
neutrino potentials never exceeds unity unlike in the case of CCM
correlations (see Fig. \ref{fig:cop}). Actually, the UCOM SRC imitate 
the effect of the FNS with form-factor cut-off of about 850 MeV.  
The two-nucleon wave function can be treated as two point-like objects 
for nucleon separations greater than about 1.5 fm. 

The effect of the SRC on the $0\nu\beta\beta$-decay 
NMEs has been referred mostly to case when the FNS is taken into 
account. It was found that Miller-Spencer SRC reduces the $0\nu\beta\beta$-decay
NMEs by 20-30\% and UCOM SRC by $\sim$5\% \cite{anatomy}. 
For better understanding of this effect we calculate the $0\nu\beta\beta$-decay of 
$^{76}Ge$, $^{100}Mo$ and $^{130}Te$ with and without consideration of the FNS. 
The 12 levels ($^{76}Ge$) and 13 levels ($^{100}Mo$ and $^{130}Te$) 
single particle model spaces are used in calculation.  The results 
are displayed in Table \ref{tab:fns}. 
The $bare$ NME was obtained in the limit of cut-off masses  
$M_{V,A}$ go to infinity and with the two-nucleon SRC switched off. 
The FNS values of $M^{0\nu}$ are determined by nucleon form factors with 
phenomenological values of $M_V$ and $M_A$. We see that the FNS reduces
$M^{0\nu}$ by 20 \%. The $0\nu\beta\beta$-decay NME is 
suppressed by about 20-30\% and 40-50\% in the cases of the
two-nucleon CCM (Argonne potential) and the phenomenological 
Miller-Spencer SRC, respectively. It is also shown that once
SRC effects are included the consideration of the nucleon 
form-factors almost does not influence the value of $M^{0\nu}$. 
It is because the FNS and the SRC effects act coherently 
on the $0\nu\beta\beta$-decay NMEs and both diminish them. 
However, the effect of the SRC is at least partially stronger 
(CCM SRC) or stronger (Miller-Spencer SRC) in comparison 
with the FNS effect.

\section{Numerical results}

The nuclear matrix elements  of the $0\nu\beta\beta$-decay 
of the experimentally interesting nuclei $A=76$, $82$, $96$, 
$100$, $116$, $128$, $130$ and $136$ are systematically 
evaluated using the QRPA and RQRPA. In the present calculations, 
we improve on the Miller-Spencer Jastrow and UCOM methods 
by engaging the SRC calculated within the
exp(S) approach with the CD-Bonn and Argonne V18 
NN interactions. This allows a consistent study of the 
$0\nu\beta\beta$-decay NMEs as for the first time
the same realistic nucleon-nucleon
force is used for the description of the pairing interactions, 
RPA ground state correlations  and the two-nucleon SRC. 

The nuclear structure calculations are performed as described
in our previous publications \cite{Rod06,Rod03a,anatomy}. Three different 
single-particle model spaces are used: small (2-3 oscillator shells), 
intermediate (3-4 oscillator shells) and large (5 oscillator shells)
model spaces (see Ref. \cite{Rod03a}).
The single-particle energies are obtained by using a
Coulomb-corrected Woods-Saxon potential \cite{bohr}. The interactions
employed are the Brueckner G-matrices which are a solution of
the Bethe-Goldstone equation with the CD-Bonn and Argonne V18
one-boson exchange potentials. The pairing two-body
interaction is fitted in the standard way and the pairing parameters 
of the BCS are adjusted to reproduce the phenomenological pairing 
gaps, extracted from the atomic mass tables. We renormalize 
the particle-particle and particle-hole channels of the G-matrix 
interaction of the nuclear Hamiltonian by introducing 
the parameters $g_{pp}$ and $g_{ph}$, respectively. We 
use $g_{ph} = 1$ throughout what allows to reproduce well 
the available data on the position of the giant Gamow-Teller 
resonance. The particle-particle strength parameter $g_{pp}$ 
of the (R)QRPA is fixed by the data on the two-neutrino double 
beta decays. 

The NME calculated within the above procedure, which 
includes three different model spaces, is denoted as the averaged 
$0\nu\beta\beta$-decay NME $\langle {M'}^{0\nu} \rangle$.
The results are presented separately for the 
CD-Bonn and Argonne interactions
and for two different values of the axial coupling constant $g_A$ 
in Table \ref{tab:res}. We confirm again that with considered 
procedure the $0\nu\beta\beta$-decay values become essentially 
independent on the size of the single-particle basis and
rather stable with respect to the possible quenching 
of the $g_A$. The NMEs obtained with the CD-Bonn NN interaction are
slightly larger as those calculated with the Argonne interaction. 
This is explained by the fact that the  CCM Argonne 
correlation function cuts out more the small $r_{12}$ part from the relative
wave function of the two-nucleons as the CCM CD-Bonn  
correlation function. The differences in NMEs due to a different
treatment of the SRC do not exceed differences between the QRPA
and the RQRPA results.

\begin{figure}[tb]
\includegraphics[width=.48\textwidth]{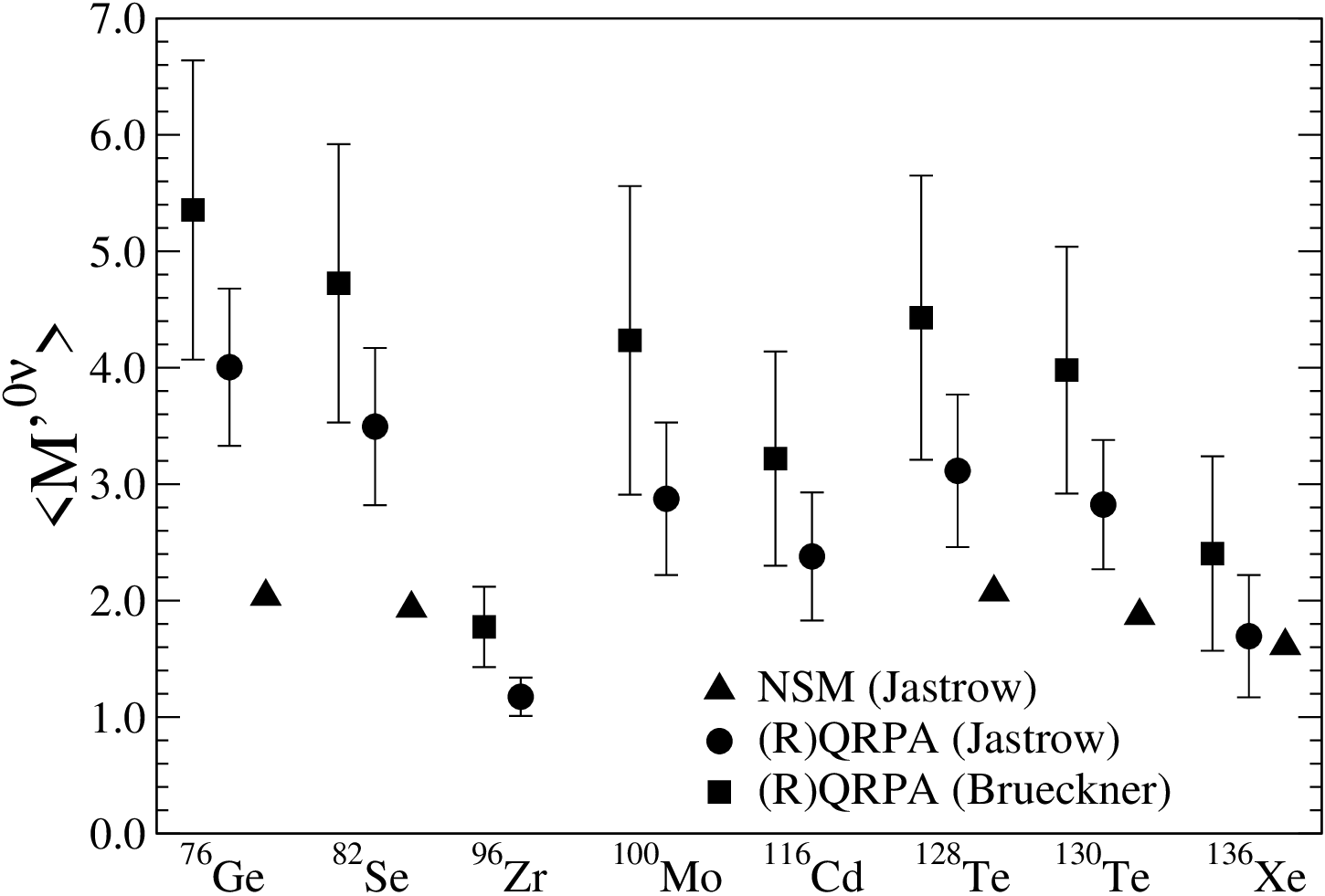}
\caption{The full ranges of $M^{'0\nu}$ with the
CCM and Miller-Spencer Jastrow treatments of the short range 
correlations. For comparison the results of a recent Large 
Scale Shell Model evaluation of $M^{'0\nu}$ that used the 
Jastrow-type treatment of short range
correlations are also shown. } \label{fig:total}
\end{figure}

In Table \ref{tab:t12} we show the calculated ranges of 
the nuclear matrix element $M^{'0\nu}$ evaluated within
the QRPA and RQRPA, with standard ($g_A = 1.254$) and
quenched ($g_A = 1.0$) axial-vector couplings and with the CCM
CD-Bonn and Argonne SRC functions. 
These ranges quantify the uncertainty in the calculated
$0\nu\beta\beta$-decay NMEs. In respect to the central
value their accuracy  is of the order  of 25\%. A significant
amount of the uncertainty is due to a
quenching of the axial-vector coupling constant $g_A$ in 
nuclear medium \cite{lisi2}. For a comparison we present also the NMEs
calculated with the phenomenological Jastrow SRC function
in Table \ref{tab:t12}. The notable differences between the
results calculated with Jastrow and CCM SRC are about of
20-30\%. Of course, the results obtained with the CCM
SRC are preferable. Unfortunately, they can not be 
directly compared to those of the complementary 
Large-Scale Shell Model (LSSM) as they have been evaluated
only with the Jastrow and UCOM SRC and for $g_A = 1.25$ \cite{LSSM3,smanat}.
It is reasonable to assume that the LSSM values (see 
Table \ref{tab:t12}) would be increased also by the
order of 20-30\%, if the CCM SRC would be
considered. Recall, that within both approaches qualitatively
the same $r_{12}$ dependence of $M^{0\nu}$ was found 
\cite{anatomy,smanat}. 
Fig. \ref{fig:total} shows our calculated ranges for $M'^{0\nu}$,
which are compared with ranges calculated with 
Miller-Spencer Jastrow function and the latest LSSM results.
Given the interest in the subject, we show the range of 
predicted half-lives corresponding to our full range of 
$M^{0\nu}$ in Table \ref{tab:t12}. 

Recently, the occupation numbers of neutron and proton valence 
orbits in the $^{76}Ge$ and  $^{76}Se$ nuclei  were measured 
by neutron and proton adding and removing transfer reactions \cite{schif1,schif2}.
In following theoretical study \cite{ocnbb} these results 
were used as a guideline 
for modification of the effective mean field energies that results 
in better description of these quantities. The calculation of the
$0\nu\beta\beta$-decay NME for $^{76}Ge$ performed with adjusted
mean field \cite{ocnbb}
and in combination with the selfconsistent RQRPA
(SRQRPA) method \cite{srpa}, which conserves the mean particle number in
correlated ground state, lead to a reduction of $M^{0\nu}$ 
by 25\% when compared to previous QRPA value. The phenomelogical
Jastrow and UCOM SRC were considered. We found that a reduction of 
20\% in the case of CCM CD-Bonn SRC. We have
\begin{eqnarray}
<M^{0\nu}> &=& 4.24(0.44),~~3.49(0.23),~~Jastrow 
\nonumber \\
           &=& 5.19(0.54),~~4.60(0.23),~~UCOM
\nonumber \\
           &=& 6.32(0.32),~~5.15(0.44),~~CCM.
\nonumber\\
\end{eqnarray}
Here, the first and second values in each line 
correspond to the QRPA with Woods-Saxon mean field 
and the SRQRPA with adjusted Woods-Saxon mean field \cite{ocnbb}
way of calculations, respectively.
In order to better understand the role of proton and neutron 
occupation numbers in the $0\nu\beta\beta$-decay calculation
further experimental and theoretical studies are needed what goes
beyond the scope of this paper. 

\section{Conclusions}

We have addressed the issue of a consistent treatment of the
short-range correlations in the context of the $0\nu\beta\beta$-decay. 
These correlations, which have origin in the short-range repulsion 
of the realistic NN interaction, are neglected in the mean-field,
the LSSM and the QRPA descriptions. Till now,  
Miller-Spencer Jastrow and the UCOM SRC
have been introduced into the involved two-body
transition matrix elements, changing two neutrons into two protons,
to achieve healing of the correlated  wave functions.
The effect of these SRC was considered as an uncertainty \cite{anatomy}.

In this article the CCM short-range correlations has been 
considered. They were obtained as a solution of the coupled cluster 
method with realistic CD-Bonn and Argonne V18 interactions. An 
analysis of the squared correlation functions,
represented by a ratio of correlated and uncorrelated 
neutrino potentials, has showed a principal differences among
the Miller-Spencer, UCOM and CCM SRC. In addition,
the importance of the effect of the FNS was studied. It was
found that both CCM SRC and the FNS effect reduces
the $0\nu\beta\beta$-decay NMEs by a comparable amount
for a considered choice of form-factor masses. The suppression
due to Miller-Spencer SRC is about twice larger when
compared to results without SRC and the FNS effect.  
 
Finally, we have improved the presently available calculations 
by performing a consistent calculation of the $0\nu\beta\beta$-decay
NMEs  in which pairing, ground-state correlations 
and the short-range correlations originate from the same realistic 
NN interaction, namely from the CD-Bonn and Argonne potentials.

\acknowledgments

Discussions with Petr Vogel are gratefully acknowledged.
We acknowledge the support of  the EU ILIAS project under the contract 
RII3-CT-2004-506222, the Deutsche Forschungsgemeinschaft 
(436 SLK 17/298), the Transregio Project TR27 "Neutrinos and Beyond" and, 
in addition, F.\v{S} was supported by  the VEGA Grant agency of
the Slovak Republic under the contract No.~1/0639/09.


\begin{table*}[htb]  
  \begin{center}  
    \caption{
Nuclear matrix elements for the $0\nu\beta\beta$-decays of $^{76}Ge$, $^{100}Mo$
and $^{136}Te$ within the QRPA. The results are presented:
i) (bare) no correlations and no nucleon form factors; 
ii) (FNS) no correlations but with nucleon form factors;
iii) (SRC) CCM Argonne and Miller-Spencer 
short-range correlations but without nucleon form factors;
iv) (FNS+SRC) correlations and nucleon form factors.
}  
\label{tab:fns}  
\renewcommand{\arraystretch}{1.2}  
\begin{tabular}{lcccccccccc}  
\hline\hline  
 Nucleus & & bare & & FNS & &  
 \multicolumn{2}{c} {SRC} & & 
 \multicolumn{2}{c} {FNS + SRC} \\ 
 \cline{7-8}  \cline{10-11}  
 & & & & & & CCM & Miller-Spencer &   & CCM & Miller-Spencer \\
\hline   
$^{76}Ge\rightarrow {^{76}Se}$ & &
 7.39 & & 6.14 & & 5.86 & 4.46 &  & 5.91 & 4.54 \\
$^{100}Mo\rightarrow {^{100}Ru}$ & &
 6.15 & & 4.75 & & 4.40 & 2.87 &  & 4.46 & 2.96 \\
$^{130}Te\rightarrow {^{130}Xe}$ & &
 5.62 & & 4.49 & & 4.22 & 2.97 & & 4.27 & 3.04 \\ \hline \hline 
\end{tabular}  
  \end{center}  
\end{table*}

\begin{table*}[htb]  
  \begin{center}  
    \caption{Averaged $0\nu\beta\beta$ nuclear matrix elements 
$\langle {M'}^{0\nu} \rangle$ and their variance $\sigma$ 
(in parenthesses) calculated within the RQRPA and
the QRPA. The pairing and residual interactions of the nuclear 
Hamiltonian and the two-nucleon short-range correlations (SRC) are
derived from the same realistic nucleon-nucleon interaction
(CD-Bonn and Argonne potentials) by exploiting the 
Brueckner-Hartree-Fock and coupled cluster  (CCM) methods. 
Three sets of single particle level schemes are used, ranging in
size from 9 to 23 orbits. 
The strength of the particle-particle interaction is adjusted 
so the experimental value of the $2\nu\beta\beta$-decay nuclear 
matrix element $M_{GT}^{exp}$ is correctly reproduced. Both
free nucleon ($g_A = 1.254$) and quenched ($g_A = 1.0$) values 
of axial-vector coupling constant are considered.
}  
\label{tab:res}  
\begin{tabular}{lccccccc}  
\hline\hline  
 Nuclear & $~g_A~~$ & $M_{GT}^{exp}$ & 
 \multicolumn{2}{c} {$\langle {M'}^{0\nu} \rangle$} & &  
 \multicolumn{2}{c} {$\langle {M'}^{0\nu} \rangle$} \\ 
 \cline{4-5}  \cline{7-8}  
transition & & & 
 \multicolumn{2}{c} {CCM CD-Bonn SRC} & &  
 \multicolumn{2}{c} {CCM Argonne SRC} \\ 
 \cline{4-5}  \cline{7-8}  
& & [MeV$^{-1}$] 
& \hspace{0.3cm} RQRPA \hspace{0.3cm} & \hspace{0.3cm} QRPA \hspace{0.3cm}  &
& \hspace{0.3cm} RQRPA \hspace{0.3cm} & \hspace{0.3cm} QRPA \hspace{0.3cm}  \\  
\hline   

$^{76}Ge\rightarrow {^{76}Se}$
    & 1.25 & $0.15$  & 5.44(0.23) & 6.32(0.32) & & 4.97(0.19) & 5.81(0.27) \\
    & 1.00 & $0.23$  & 4.62(0.22) & 5.16(0.25) & & 4.21(0.14) & 4.77(0.20) \\

$^{82}Se\rightarrow {^{82}Kr}$     
    & 1.25 & $0.10$  & 4.86(0.20) & 5.65(0.27) & & 4.44(0.19) & 5.19(0.24) \\ 
    & 1.00 & $0.16$  & 3.93(0.15) & 4.48(0.20) & & 3.67(0.14) & 4.19(0.18) \\

$^{96}Zr\rightarrow {^{96}Mo}$     
    & 1.25 & $0.11$  & 2.01(0.20) & 2.09(0.03) & & 1.84(0.16) & 1.90(0.09) \\
    & 1.00 & $0.17$  & 1.72(0.15) & 1.93(0.11) & & 1.55(0.12) & 1.74(0.11) \\

$^{100}Mo\rightarrow {^{100}Ru}$     
    & 1.25 & $0.22$  & 4.28(0.28) & 5.25(0.31) & & 3.85(0.31) & 4.75(0.33) \\
    & 1.00 & $0.34$  & 3.44(0.19) & 4.07(0.22) & & 3.14(0.23) & 3.69(0.25) \\

$^{116}Cd\rightarrow {^{116}Sn}$     
    & 1.25 & $0.12$  & 3.41(0.24) & 3.99(0.15) & & 3.06(0.22) & 3.54(0.27) \\
    & 1.00 & $0.19$  & 2.68(0.19) & 3.03(0.19) & & 2.47(0.17) & 2.74(0.21) \\
$^{128}Te\rightarrow {^{128}Xe}$     
    & 1.25 & $0.034$ & 4.82(0.15) & 5.49(0.16) & & 4.32(0.16) & 4.93(0.16) \\
    & 1.00 & $0.053$ & 3.67(0.11) & 4.16(0.12) & & 3.32(0.11) & 3.77(0.12) \\

$^{130}Te\rightarrow {^{130}Xe}$  
    & 1.25 & $0.036$ & 4.40(0.13) & 4.92(0.12) & & 3.91(0.14) & 4.37(0.14) \\
    & 1.00 & $0.056$ & 3.38(0.08) & 3.77(0.07) & & 3.02(0.10) & 3.38(0.10) \\

$^{136}Xe\rightarrow {^{136}Ba}$ 
    & 1.25 & $0.030$ & 2.89(0.17) & 3.11(0.13) & & 2.59(0.16) & 2.78(0.13) \\
    & 1.00 & $0.045$ & 2.26(0.11) & 2.42(0.08) & & 2.03(0.10) & 2.17(0.09) \\
    & 1.25 & 0       & 2.53(0.17) & 2.73(0.13) & & 2.25(0.16) & 2.43(0.13) \\
    & 1.00 & 0       & 1.87(0.11) & 2.01(0.08) & & 1.67(0.10) & 1.80(0.09) \\

\hline\hline  
\end{tabular}  
  \end{center}  
\end{table*}

\begin{table*}[htb]
  \begin{center}
    \caption{\label{tab:t12}
The calculated ranges of the nuclear matrix element
    $M^{'0\nu}$ evaluated within
the QRPA and RQRPA, with standard ($g_A = 1.254$) and
quenched ($g_A = 1.0$) axial-vector couplings and with  coupled cluster 
method (CCM) CD-Bonn and Argonne short-range correlation (SRC) functions. 
Column 4 contains the ranges of $M^{'0\nu}$ with the phenomenological 
Miller-Spencer Jastrow treatment of short range correlations, while
column 6 shows the CCM SRC-based results.
For comparison the recent results of a large scale shell model (LSSM)
evaluation of ${M'}^{0\nu}$ \protect\cite{LSSM3}
that used the Miller-Spencer Jastrow
SRC and $g_A = 1.25$ are given in column 2. 
However, they have to be scaled by factor $(1.1~fm/1.2~fm)$ as
different value of $r_0$ ($R_{nucl} = r_0 A^{1/3}$) was considered.
Columns 3, 5 and 7 give the  $0\nu\beta\beta$-decay half-lives or
half-life ranges corresponding to values of the matrix-elements 
in columns 2, 4 and 6 for $<m_{\beta\beta}>=50$~meV.
}
\begin{tabular}{lcccccccc}
\hline\hline
 Nucleus & \multicolumn{2}{c}{LSSM (Jastrow SRC)} & &
           \multicolumn{2}{c}{(R)QRPA (Jastrow SRC)} & &
           \multicolumn{2}{c}{(R)QRPA (CCM SRC)}\\ 
           \cline{2-3} \cline{5-6} \cline{8-9}
 & $M^{0\nu}$ & $T^{0\nu}_{1/2}$ ($\langle m_{\beta\beta} \rangle$ = 50 meV) &
 & $M^{'0\nu}$ & $T^{0\nu}_{1/2}$ ($\langle m_{\beta\beta} \rangle$ = 50 meV) &
 & $M^{'0\nu}$ & $T^{0\nu}_{1/2}$ ($\langle m_{\beta\beta} \rangle$ = 50 meV) \\\hline
$^{76}Ge$
  &  $2.22$ & $3.18\times 10^{27}$ &
  &  $(3.33,4.68)$ & $(6.01,11.9)\times 10^{26}$ &
  &  $(4.07,6.64)$ & $(2.99,7.95)\times 10^{26}$ \\
$^{82}Se$
  &  $2.11$ & $7.93\times 10^{26}$ &
  &  $(2.82,4.17)$ & $(1.71,3.73)\times 10^{26}$ &
  &  $(3.53,5.92)$ & $(0.85,2.38)\times 10^{26}$ \\
$^{96}Zr$
  &  &  &
  &  $(1.01,1.34)$ & $(7.90,13.9)\times 10^{26}$ &
  &  $(1.43,2.12)$ & $(3.16,6.94)\times 10^{26}$ \\
$^{100}Mo$
  &  &  &
  &  $(2.22,3.53)$ & $(1.46,3.70)\times 10^{26}$ &
  &  $(2.91,5.56)$ & $(0.59,2.15)\times 10^{26}$ \\
$^{116}Cd$
  &  &  &
  &  $(1.83,2.93)$ & $(1.95,5.01)\times 10^{26}$ &
  &  $(2.30,4.14)$ & $(0.98,3.17)\times 10^{26}$ \\
$^{128}Te$
  &  $2.26$ & $1.10\times 10^{28}$ &
  &  $(2.46,3.77)$ & $(3.33,7.81)\times 10^{27}$ &
  &  $(3.21,5.65)$ & $(1.48,4.59)\times 10^{27}$ \\
$^{130}Te$
  &  $2.04$ & $5.39\times 10^{26}$ &
  &  $(2.27,3.38)$ & $(1.65,3.66)\times 10^{26}$ &
  &  $(2.92,5.04)$ & $(7.42,2.21)\times 10^{26}$ \\
$^{136}Xe$
  &  $1.70$ & $6.79\times 10^{26}$ &
  &  $(1.17,2.22)$ & $(3.59,12.9)\times 10^{26}$ &
  &  $(1.57,3.24)$ & $(1.68,7.17)\times 10^{26}$ \\
\hline\hline
\end{tabular}
  \end{center}
\end{table*}

\end{document}